\documentstyle[prl,aps,epsf,epsfig,floats]{revtex}
\begin{document}
\twocolumn[\hsize\textwidth\columnwidth\hsize\csname@twocolumnfalse%
\endcsname
\draft
%

\title{\bf Photoemission Evidence for a Remnant Fermi Surface 
and {\it d}-Wave-Like Dispersion in Insulating
Ca$_{2}$€Cu€O$_{2}$€Cl$_{2}$€}

\author{F.~Ronning$^{(1)}$, C.~Kim$^{(1)}$, D.L.~Feng$^{(1)}$, 
D.S.~Marshall$^{(1)}$, 
A.G.~Loeser$^{(1)}$, L.L.~Miller$^{(2)}$, J.N.~Eckstein$^{(3)}$, 
I.~Bozovic$^{(3)}$, Z.-X.~Shen$^{(1)}$}

\address{$^{(1)}$Department of Physics and Stanford 
Synchrotron Radiation Laboratory, Stanford University,
Stanford, CA 94305-4045}

\address{$^{(2)}$ Department of Physics, Iowa State University, Ames, IA
50011}

\address{$^{(3)}$ Ginzton Research Center, Varian Associates, Palo Alto,
CA 94304}

\maketitle


\begin{abstract}

An angle resolved photoemission study on 
Ca$_{2}$€Cu€O$_{2}$€Cl$_{2}$€, a parent compound of high T$_{c}$ 
superconductors is reported.  Analysis of the electron occupation 
probability, n({\it k}), from the spectra shows a steep drop in 
spectral intensity across a contour that is close to the Fermi surface 
predicted by the band calculation.  This analysis reveals a Fermi 
surface remnant even though Ca$_{2}$€Cu€O$_{2}$€Cl$_{2}$€ is a Mott 
insulator.  The lowest energy peak exhibits a dispersion with 
approximately the $|$cos($\it k$$_{x}$a)-cos($\it k$$_{y}$a)$|$ form along this 
remnant Fermi surface.  Together with the data from Dy doped 
Bi$_{2}$€Sr$_{2}$€CaCu$_{2}$€O$_{8 + \delta}$€ these results suggest 
that this {\it d}-wave like dispersion of the insulator is the 
underlying reason for the pseudo gap in the underdoped regime.

\vspace{0.2in}
\end{abstract}

]
\narrowtext

\section{Introduction}
Introduction: A consensus on the d$_{x^{2}-y^{2}}$ pairing state and the 
basic phenomenology of the anisotropic normal state gap (pseudo gap) 
in high-T$_{c}$ superconductivity has been established~\cite{Levi} , 
partially on the basis of angle-resolved photoemission spectroscopy 
(ARPES) experiments \cite{Shen} \cite{Loeser} \cite{Marshall} 
\cite{Ding}, in which two energy scales have been identified in the 
pseudo gap, a leading-edge shift of 20-25 meV and a high-energy hump 
at 100-200 meV.  \cite{Marshall} Both of these features have an 
angular dependence consistent with a {\it d}-wave gap.  For simplicity 
in the discussion below, we refer to these as low- and high-energy 
pseudo gaps, respectively, in analogy to the analysis of other data.  
\cite{Battlog} The evolution of these two pseudo gaps as a function of 
doping are correlated \cite{White}, but the microscopic origin of the 
pseudo gap and its doping dependence are still unestablished.  
Theoretical ideas of the pseudo gap range from pre-formed pairs or 
pair fluctuation \cite{Kivelson} and damped spin density wave (SDW) 
\cite{Schmallian} to the evidence of the resonating valence bond (RVB) 
singlet formation and spin-charge separation \cite{Tanamoto} 
\cite{Wen} \cite{Laughlin}.  

To further differentiate these ideas, it 
is important to understand how the pseudo gap evolves as the doping is 
lowered and the system becomes an insulator.  We present experimental 
data from the insulating analog of the superconductor 
La$_{2-x}$€Sr$_{x}$€CuO$_{4}$€, Ca$_{2}$€Cu€O$_{2}$€Cl$_{2}$€ which 
suggest that the high energy pseudo gap is a remnant property of the 
insulator that evolves continuously with doping, as first pointed out 
by Laughlin.  \cite{Laughlin} The Compound 
Ca$_{2}$€Cu€O$_{2}$€Cl$_{2}$€, a half-filled Mott insulator, has the 
crystal structure of La$_{2}$€CuO$_{4}$€ \cite{Miller} and it can be 
doped by replacing Ca with Na or K to become a high-temperature 
superconductor.  \cite{Hiroi} As with the case of 
Sr$_{2}$€Cu€O$_{2}$€Cl$_{2}$€, Ca$_{2}$€Cu€O$_{2}$€Cl$_{2}$€ has a 
much better surface property than La$_{2}$€CuO$_{4}$€ and thus is 
better suited for ARPES experiments.  \cite{Experiment} Although the 
data from Ca$_{2}$€Cu€O$_{2}$€Cl$_{2}$€ are consistent with earlier 
results from Sr$_{2}$€Cu€O$_{2}$€Cl$_{2}$€, \cite{Wells} \cite{LaRosa} 
\cite{Kim} the improved spectral quality obtainable from this material 
allows us to establish that: (I) The Fermi surface, which is destroyed 
by the strong Coulomb interactions, left a remnant in this insulator 
with a volume and shape similar to what one expects if the strong 
electron correlation in this system is turned off; (II) The strong 
correlation effect deforms this otherwise iso-energetic contour (the 
non-interacting Fermi surface) into a form that matches the 
$|$cos($\it k$$_{x}$a)-cos($\it k$$_{y}$a)$|$ function very well, but with a very high 
energy scale of 320 meV.  Thus, a Œ{\it d}-wave¹ like dispersive 
behavior exists even in the insulator.

Comparison with data from underdoped Bi$_{2}$€Sr$_{2}$€CaCu$_{2}$€O$_{8+\delta}$€ (Bi2212) with T$_{c}$'s 
of 0, 25 and 65 K indicates that the high energy {\it d}-wave like 
pseudo gap in the underdoped regime originates from the {\it d}-wave 
like dispersion in the insulator.  Once doped to a metal, the chemical 
potential drops to the maximum of this {\it d}-wave like function, but 
the dispersion relation retains its qualitative shape, albeit the 
magnitude decreases with doping.  Thus, only the states near the {\it 
d}-wave node touch the Fermi level and form small segments of the 
Fermi surface, with the rest of Fermi surface gapped.  In this way, 
the {\it d}-wave high energy pseudo gap in the underdoped regime is 
naturally connected to the properties of the insulator.  Since the 
high energy pseudo gap correlates with the low energy pseudo gap which 
is likely to be related to superconductivity \cite{Loeser} 
\cite{Marshall} \cite{Ding} \cite{White} \cite{ZX}, it is likely 
that the same physics that controls the {\it d}-wave dispersion in the 
insulator is responsible for the {\it d}-wave like normal state pseudo 
gap and the superconducting gap in the doped superconductors.

\section{Methodology}
To investigate the strong correlation effect, we contrast 
our experimental data with the conventional results for the case when 
the correlation effects are neglected.  We can obtain the occupation 
probability, n({\it k}), by integrating A({\it k},$\omega$) obtained 
by ARPES, over energy.  \cite{Hufner} Experimentally A({\it 
k},$\omega$) can not be integrated over all energies due to 
contributions from secondary electrons and other electronic states.  
Instead an energy window for integration must be chosen, and the 
resulting quantity we define as the {\it relative} n({\it k}).  Fortunately, 
the features we are interested in are clearly distinguishable from any 
other contributions.  We note that n({\it k}) is a ground state 
property, and hence is different from the integration of the 
single-particle spectral weight, A({\it k},$\omega$), over energy.  
However, under the sudden approximation integration of A({\it 
k},$\omega$) as measured by ARPES gives n({\it k}).  \cite{Hufner} We 
then use the drop of the relative n({\it k}) to determine the Fermi 
surface as illustrated in Fig.  1.  For a metal with non-interacting 
electrons, the electron states are filled up to the Fermi momentum, 
{\it k$_{F}$}, and the n({\it k}) shows a sudden drop(Fig.  1A).  As 
more electrons are added, the electron states are eventually filled 
and the system becomes an insulator with no drop in n({\it k}) (Fig.  
1B).  Therefore, the drop in n({\it k}) characterizes the Fermi 
surface of a metal with non-interacting electrons.  When correlation 
increases, n({\it k}) begins to deform (Fig.  1C), although there is 
still some discontinuity at {\it k$_{F}$} when the correlation is 
moderate.  Note that the electrons that used to occupy states below 
{\it k$_{F}$} have moved to the states that were unoccupied.  For a 
non-Fermi liquid with very strong correlation, n({\it k}) drops 
smoothly without a discontinuity(panel D).  Several theoretical 
calculations using very different models have found that n({\it k}) of 
the interacting system mimics that of the non-interacting system, even 
when the material is fully gapped \cite{Putikka} \cite{Bulut} 
\cite{Eder} \cite{Dagatto}.  Hence we can recover the remnant of a 
Fermi surface or an underlying Fermi surface by following the contour 
of steepest descent of n({\it k}) even when correlation is strong 
enough that the system becomes a Mott insulator.  \cite{key} The 
volume obtained by this procedure is consistent with half-filling as 
expected in a Mott insulator.

We apply this method to determine the Fermi level crossing of a real 
system.  The traditional way (Fig.  2A) is shown for the ARPES spectra 
on the (0,0) to $(\pi,\pi)$ cut taken from Bi2212 which is metallic.  
As we move from (0,0) toward $(\pi,\pi)$, the peak disperses to the 
Fermi level, E$_{F}$.  As the peak reaches E$_{F}$ and passes it, it 
begins to lose spectral weight (this again is {\it k}$_{F}$).  
Alternatively, we simply integrate the spectral function from 0.6 eV 
to -0.1 eV relative to the E$_{F}$, and the resulting relative n({\it 
k}) is plotted in Fig.  2C. We can now define {\it k}$_{F}$ as the 
point of steepest descent in the relative n({\it k}).  The same 
conclusion can be drawn here independent of the method we use.  Note 
that the n({\it k}) also drops as we approach (0,0); this is a 
photoemission artifact, because the photoemission cross-section of the 
d$_{x^{2}-y^{2}}$ orbital vanishes due to symmetry.

We can show that the n({\it k}) procedure is still valid for strongly 
correlated systems with gapped Fermi surface by presenting ARPES 
spectra on ferromagnetic La$_{3-x}$€Sr$_{x}$€Mn$_{2}$€O$_{7}$€ on the 
$(\pi,0)$ to $(\pi,\pi)$ cut (Fig.  2B). \cite{Dessau}  It shows a dispersive 
feature initially moving toward E$_{F}$ and then pulling slightly back 
away from E$_{F}$ around $(\pi,0.27\pi)$, but never reaching the 
E$_{F}$.  However, the feature suddenly loses its spectral weight when 
it crosses $(\pi,0.27\pi)$ as if it crosses the Fermi surface as shown 
in panel D.  Furthermore, the Fermi surface determined by a local 
density approximation calculation coincides with the Fermi surface 
determined by the n({\it k}) despite the spectra of this ferromagnetic 
metallic state material having a significant gap.  Thus, the 
underlying Fermi surface can survive a strong interaction, and the 
n({\it k}) method is effective in identifying it even when the peak 
does not disperse across E$_{F}$. 

\section{Experimental results}
The low-energy feature along the (0,0) to $(\pi,\pi)$ cut on 
Ca$_{2}$€Cu€O$_{2}$€Cl$_{2}$€ (Fig.  3A) has the same origin as the 
lo- energy peak seen in Bi2212, the Zhang-Rice singlet on the 
Cu€O$_{2}$€ plane.  As {\it k} increases from (0,0) toward 
$(\pi,\pi)$, the peak moves to lower energy and subsequently pulls 
back to higher energy as it crosses $(\pi /2,\pi /2)$.  Its spectral 
weight increases as it moves away from the (0,0) point for the reason 
described earlier, and then drops as it crosses $(0.43\pi ,0.43\pi)$.  
These changes along the (0,0) to $(\pi,\pi)$ cut are consistent with 
the earlier reports on Sr$_{2}$€Cu€O$_{2}$€Cl$_{2}$€.  \cite{Wells} 
\cite{LaRosa} \cite{Kim} Similar to the drop of n({\it k}) across the 
Fermi surface seen in Bi2212, Ca$_{2}$€Cu€O$_{2}$€Cl$_{2}$€ also shows 
that the intensity of the peak n({\it k}) drops as if there is a 
crossing of E$_{F}$ even though the material is an insulator.  The 
intensity along the (0,0) to $(\pi,0)$ cut (Fig.  3B) goes through a 
maximum around $(2\pi /3,0)$ as in Sr$_{2}$€Cu€O$_{2}$€Cl$_{2}$€.  
This behavior is also seen in superconducting cuprates.  \cite{Dan} 
Earlier works on Sr$_{2}$€Cu€O$_{2}$€Cl$_{2}$€ show the spectral 
weight along the $(\pi,0)$ to $(\pi,\pi)$ cut is strongly suppressed.  
However, for Ca$_{2}$€Cu€O$_{2}$€Cl$_{2}$€, the improved spectral 
quality allows us to clearly observe the spectral weight drop along 
the $(\pi,0)$ to $(\pi,\pi)$ cut (Fig.  3C).  \cite{generally} Note 
that the spectral weight drops as we move toward the $(\pi,\pi)$ 
point, which we attribute to the crossing of a remnant Fermi surface.  
We also show another cut (Fig.  3D) which exhibits essentially the 
same behavior.  The relative n({\it k})¹s of the cuts are summarized 
in Fig.  3E in arbitrary units.  The relative n({\it k}) here and in 
fig.  4 were obtained by integrating from 0.5 eV to -0.2 eV relative 
to the peak position at $(\pi /2,\pi /2)$.  All of the n({\it k})¹s 
show a drop (after the maximum) as we cross the remnant Fermi surface.  
Here we emphasize that we are using the same method as we do for 
metals, where the identification of a Fermi surface is convincing.

The remnant Fermi surface can be identified in the contour plot of 
n({\it k}) of Ca$_{2}$€Cu€O$_{2}$€Cl$_{2}$€ (Fig.  4A).  The little 
crosses in the figure denote the {\it k}-space points where spectra 
were taken.  The data points here and in Fig.  4C have been reflected 
about the line {\it k}$_{y}$€={\it k}$_{x}$€ to better illustrate the 
remnant Fermi surface.  Again, it should be emphasized that the 
suppressed n({\it k}) near (0,0) comes from the vanishing 
photoemission cross section due to the d$_{x^{2}-y^{2}}$ orbital symmetry 
rather than a remnant Fermi surface crossing.  For the same reason, 
the photon polarization suppresses the overall spectral weight along 
the (0,0) to $(\pi,\pi)$ line as compared with the (0,0) to $(\pi,0)$ 
line, with a monatonic change between the two directions.  In fig.  4B 
we present the relative n({\it k}) of an optimally doped Bi2212 sample 
in the normal state.  In this case the identification of the Fermi 
surface is unambiguous, but the same matrix element effects that were 
seen in the insulator can be seen in the metallic sample as well.  
However, for both samples, the drop in n({\it k}) near the diagonal 
line connecting $(\pi,0)$ and $(0,\pi)$ can not be explained by the 
photoemission cross section.  In the metallic case, the Fermi surface 
is clearly identified (the white-hashed region in Fig.  4B).  For the 
insulator, the drop is approximately where band theory predicts the 
Fermi surface.  \cite{Novikov} Therefore, we attribute the behavior in 
the insulator to a remnant of the Fermi surface that existed in the 
metal.  The similarity of the results in the insulator and the metal 
makes the identification of the remnant Fermi surface unambiguous.  
The white hashed area in Fig.  4A represents the area where the 
remnant Fermi surface may reside as determined by the relative n({\it 
k}).  Although there is some uncertainty in the detailed shape of this 
remnant Fermi surface, this does not affect the discussion and the 
conclusions drawn below.  The relative n({\it k}) we presented is a 
very robust feature.  In metallic samples with partially gapped Fermi 
surfaces, underlying Fermi surfaces have also been identified in the 
gapped region.  \cite{Dessau} \cite{HDing} This effect is similar to 
what we report here in the insulator.  The remnant Fermi surface in 
the underdoped Bi2212 was also identified at similar locations to the 
n({\it k}) drop in these materials with a different criteria of 
minimum gap locus.  \cite{HDing} Calculations also show the Fermi 
surface defined by n({\it k}) is robust in the presence of strong 
correlation.  \cite{Putikka} \cite{Eder} \cite{Dagatto} Given that 
there is a remnant Fermi Surface as shown by the white hashed lines in 
Fig.  4, A and C, the observed energy dispersion along this line has 
to stem from the strong electron correlation.  In other words, the 
electron correlation disperses the otherwise iso-energetic contour of 
the remnant Fermi surface.  This dispersion is consistent with the 
non-trivial {\it d}-wave $|$cos($\it k$$_{x}$a)-cos($\it k$$_{y}$a)$|$ form.  
\cite{Tanamoto} \cite{Wen} \cite{Laughlin} These results also support 
our identification of the remnant Fermi surface in a Mott insulator.

Fig. 4C plots the energy contour of the peak position of the lowest 
energy feature of Ca$_{2}$€Cu€O$_{2}$€Cl$_{2}$€ referenced to the 
energy of $(\pi /2,\pi /2)$ peak.  The hashed area indicates the 
remnant Fermi surface determined in fig.  4A. The 'Fermi surface' is 
no longer a constant energy contour as it would be in the 
non-interacting case.  Instead it disperses as much as the total 
dispersion width of the system.  In Fig.  4D we plot the dispersion at 
different points on the remnant Fermi surface referenced to the lowest 
energy state at $(\pi /2,\pi /2)$.  The dispersion of the peaks along the 
Fermi surface is plotted against $|$cos($\it k$$_{x}$a)-cos($\it k$$_{y}$a)$|$.  The 
straight line shows the {\it d}-wave dispersion function at the 'Fermi 
surface' with a {\it d}-wave energy gap.  The figure in the inset 
presents the same data in a more illustrative fashion.  On a line 
drawn from the center of the Brillouin zone to any point either 
experimental (blue) or theoretical (red), the distance from this point 
to the intersection of the line with the antiferromagnetic Brillouin 
zone boundary gives the value of the 'gap' at the {\it k}-point of interest.  
The red line is for a {\it d}-wave dispersion along $(\pi,0)$ to 
$(0,\pi)$.  The good agreement  \cite{details} is achieved without the need for 
free parameters.  This {\it d}-wave like dispersion can only be 
attributed to the many-body effect.  The relative energy difference 
between the energy at $(\pi /2,\pi /2)$ and $(\pi,0)$ has been 
referred to as a Œgap¹ \cite{Laughlin}, which we follow.

This gap differs from the usual optical Mott gap (Fig.  5) and may 
correspond to the momentum dependent gap once the system is doped.  
This gap monotonically increases when we move away from $(\pi /2,\pi 
/2)$ as also reported earlier.  \cite{Wells} As well as summarizing 
the data presented, Fig.  5 also shows the intriguing similarity 
between the data from the insulator and a slightly overdoped {\it 
d}-wave superconductor(Bi2212), and thus gives the reason for 
comparing the dispersion along the remnant Fermi surface with the 
$|$cos($\it k$$_{x}$a)-cos($\it k$$_{y}$a)$|$ form.  In the superconducting case, 
n({\it k}) helps determine the Fermi surface.  The anisotropic gapping 
of this surface below T$_{c}$ reveals the {\it d}-wave nature of the 
gap.  In the insulator, n({\it k}) helps determine the remnant Fermi 
surface.  The {\it k}-dependent modulation along this surface reveals 
the {\it d}-wave like dispersion.  Whether this similarity between the 
insulator and the doped superconductor is a reflection of some 
underlying symmetry principle is a question which needs to be 
investigated.  \cite{Zhang}

The above analysis is possible only because we now observe the remnant 
Fermi surface.  Although the dispersion for 
Sr$_{2}$€Cu€O$_{2}$€Cl$_{2}$€ was similar to the present case, the 
earlier results did not address the issue of a remnant Fermi surface 
because the smaller photoemission cross section along the $(\pi,0)$ to 
$(\pi,\pi)$ cut prevented this identification.  Therefore the analysis 
shown above was not possible.  With only the energy contour 
information (such as in Fig 4C), it is plausible to think that the 
Fermi surface evolves to a small circle around the $(\pi /2,\pi /2)$ 
point.  \cite{Chubukov} However, with the favorable photoemission 
cross section, the results from Ca$_{2}$€Cu€O$_{2}$€Cl$_{2}$€ show 
that the Fermi surface leaves a clear remnant, although it may be 
broadened and weakened.  Therefore, the energy dispersion along the 
original Fermi surface of a non-interacting system is due to the 
opening of an anisotropic 'gap' along the same remnant Fermi surface.

The same analysis is shown in Fig.  6 for Bi2212 with different Dy 
dopings together with Ca$_{2}$€Cu€O$_{2}$€Cl$_{2}$€ results.  The 
corresponding doping level and T$_{c}$ as a function of Dy 
concentration are also shown.  The energy for 
Ca$_{2}$€Cu€O$_{2}$€Cl$_{2}$€ is referenced to the peak position at 
$(\pi /2,\pi /2)$ and that for Dy doped Bi2212 is to E$_{F}$.  
However, the two energies essentially refer to the same energy since 
the peak on the (0,0) to $(\pi,\pi)$ cut for all Bi2212 samples 
reaches the Fermi level.  Note that the gaps for Dy doped Bi2212 data 
also follow a function that is qualitatively similar to the {\it 
d}-wave function with reduced gap sizes as shown with the $(\pi,0)$ 
spectra in Fig.  6B. This result suggests that the {\it d}-wave gap 
originating in the insulator continuously evolves with doping, but 
retains its anisotropy as a function of momentum and that the high 
energy pseudo gap in the underdoped regime is the same gap as the {\it 
d}-wave gap seen in the insulator as discussed above.  Of course, the 
high energy pseudo gap in the underdoped regime is smaller than the 
gap in the insulator.  In a sense, the doped regime is a Œdiluted¹ 
version of the insulator, with the gap getting smaller with increasing 
doping.  The two extremes of this evolution are illustrated in the 
quasiparticle dispersions shown in Fig.  5.  The insulator shows a 
large {\it d}-wave like dispersion along the remnant Fermi surface.  
In the overdoped case, no gap is seen in the normal state along an 
almost identical curve in {\it k}-space; however, a {\it d}-wave gap 
is observed in the superconducting state.  Although their sizes vary, 
the {\it d}-wave superconducting gap, and the {\it d}-wave 'gap' of 
the insulator have the same non-trivial form, and are thus likely to 
stem from the same underlying mechanism.

\section{Discussion}
We do not know the full implications of the data we report, but can offer 
the following possibilities.  First, we compare the experimental 
dispersion with a simple spin-density wave picture. Starting with the 
Hubbard model 

H = $\Sigma$$_{k\sigma}$€ $\epsilon$$_{k}$€ 
c$^{\dagger}$ $_{k\sigma}$€c$_{k\sigma}$€ + U$\Sigma$$_{i}$€ 
n$_{i\uparrow}$€n$_{i\downarrow}$€

with 

$\epsilon$$_{k}$€ = -2t(cos($\it k$$_{x}$a)+cos($\it k$$_{y}$a)) - 
4t'(cos($\it k$$_{x}$a)cos($\it k$$_{y}$a)) - 2t"(cos(2k$_{x}$a)+cos(2k$_{y}$a))
  
and adding a SDW picture, the following dispersion relation will be found

E$_{k\pm}$€ $\approx$ -4t'(cos($\it k$$_{x}$a)cos($\it k$$_{y}$a)) - 
2t"(cos(2k$_{x}$a)+cos(2k$_{y}$a)) $\pm$ [U/2 + 
J(cos($\it k$$_{x}$a)+cos($\it k$$_{y}$a))$^{2}$€]

with J=t$^{2}$€/U.  With realistic values for t' and t", and an experimental 
value for J of -0.12 eV, 0.08 eV, and 0.125 eV respectively, \cite{Kim} we find 
that the experimental dispersion deviates significantly from this mean 
field result giving a bandwidth of 1.1 eV.  It is crucial to note the 
observed isotropic dispersion around the $(\pi /2,\pi /2)$ point, with almost 
identical dispersions from $(\pi /2,\pi /2)$ to (0,0) and from $(\pi /2,\pi /2)$ to 
$(\pi,0)$.  This result is unlikely to be a coincidence of the parameters t', 
t", and J as suggested by the SDW picture above.

We now compare the data with numerical calculations that, unlike the 
mean field SDW picture, appropriately accounts for the dynamics.  
Being mainly concerned with the dispersion relation, we concentrate 
our discussion on the t-J model as more extensive literature exists 
and as J can be independently measured.  \cite{Dagatto} Qualitatively, 
the same conclusion is expected for the Hubbard model \cite{Preuss}, 
which has the added advantage of yielding n({\it k}), but has more 
uncertainty in the parameter U. Although the t-J model correctly 
predicts \cite{Dagatto} the dispersion along (0,0) to $(\pi /2,\pi 
/2)$ quantitatively \cite{Wells}, with the band width along this 
direction solely determined by J, it incorrectly predicts the energy 
of $(\pi,0)$ to be nearly degenerate to $(\pi /2,\pi /2)$.  This is a 
serious deficiency of the t-J model, because the evolution of the 
$(\pi,0)$ feature is crucial to understand the {\it d}-wave-like 
pseudo gap.  The inclusion of the next nearest neighbor hoppings of t' 
and t" can resolve this problem.  \cite{Preuss} \cite{REder} 
\cite{Tohyama} In fact, the t-t'-t"-J model can account for both the 
dispersion and lineshape evolution over all doping levels, which is a 
remarkable success of this model.  \cite{Kim} \cite{REder} With a J/t 
ratio in the realistic range of 0.2 to 0.6, the t-t'-t"-J model shows 
that the dispersion from $(\pi /2,\pi /2)$ to (0,0) and to $(\pi,0)$ 
are equal and scaled by J. \cite{private} This result supports the 
notion that the isotropic dispersion is controlled by a single 
parameter, J, as stressed by Laughlin.  \cite{Laughlin}

The above discussion indicates that we have a model, when solved by 
Monte Carlo or exact diagonalization, that can account for the data, 
but what does the data fitting the non-trivial 
$|$cos($\it k$$_{x}$a)-cos($\it k$$_{y}$a)$|$ function so well mean?  As pointed out 
\cite{Kim}, the key to the inclusion of t' and t" is that the 
additional hole mobility destabilizes the one-hole Ne\'{e}€l state with the 
hole at $(\pi,0)$ and makes the system with one-hole move closer to a 
spin liquid state rather than to a Ne\'{e}€l state that is stable in the 
t-J model.  This point is relevant to some early literature of the 
resonating valence bond(RVB) state \cite{Anderson} \cite{SAKivelson}.  
Anderson conjectured that the ground state of the insulator at half 
filling is a RVB spin liquid state.  \cite{Anderson} This idea was 
extended in the context of a mean field approach to the t-J model that 
yields a {\it d}-wave RVB or flux phase solution.  \cite{Tanamoto} The 
mean-field solution also predicts a phase diagram similar to what is 
now known about the cuprates, with the {\it d}-wave like spin gap in 
the underdoped regime being the most successful example.  The problem 
with the mean-field solution of the t-J model is that it does not 
agree with exact numerical calculation results \cite{Dagatto}, and the 
half-filled state was found by neutron scattering to have long range 
order.  \cite{Chakravarty} If these numerical calculations are right 
then the {\it d}-wave RVB is not the right solution of the t-J model.  
However, the {\it d}-wave RVB like state may still be a reasonable way 
to think about the experimental data that describes the situation of 
the spin state near a hole.  \cite{Eskes} It is just that one has to 
start with a model where the single hole Ne\'{e}€l state is destabilized, 
as in the t-t'-t"-J model.  We leave this open question as a challenge 
to theory.

The presence of {\it d}-wave like dispersion along the remnant Fermi 
surface shows that the high energy pseudo gap is a remnant of the {\it 
d}-wave 'gap' seen in the insulator.  The details of the evolution of 
this gap, and its connection to the low energy pseudo gap (which is 
likely due to pairing fluctuations) as well as the superconducting gap 
is unclear at the moment.  However, we believe that there has to be a 
connection between these gaps of the similar 
$|$cos($\it k$$_{x}$a)-cos($\it k$$_{y}$a)$|$ form, as their presence is correlated 
with each other.  \cite{White}

\pagebreak

\begin{figure}
\leavevmode \epsfxsize=\columnwidth
\centerline{\epsffile{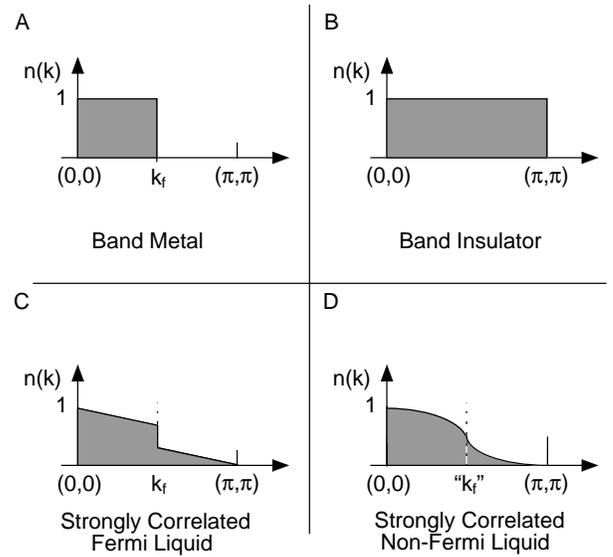}}
\vspace{0.2in}
\caption{Illustration of the Fermi surface determination.  (A) The 
case for band metal.  Electrons occupy states only up to a certain 
momentum, showing a sharp drop in n({\it k}).  (B) Band insulator 
case.  Electrons occupy all possible states and do not show a drop in 
n({\it k}).  (C) Fermi liquid with electron correlation.  Note that 
electrons that used to occupy the states below {\it k}$_{F}$ have 
moved above {\it k}$_{F}$.  However, it still shows a discontinuity at 
{\it k}$_{F}$.  (D) For a strongly correlated non-Fermi liquid n({\it 
k}) does not show discontinuity, yet there exists n({\it k}) drop 
showing the remnant Fermi surface.}
\label{Fig1}
\end{figure}

\begin{figure}
\leavevmode \epsfxsize=\columnwidth
\centerline{\epsffile{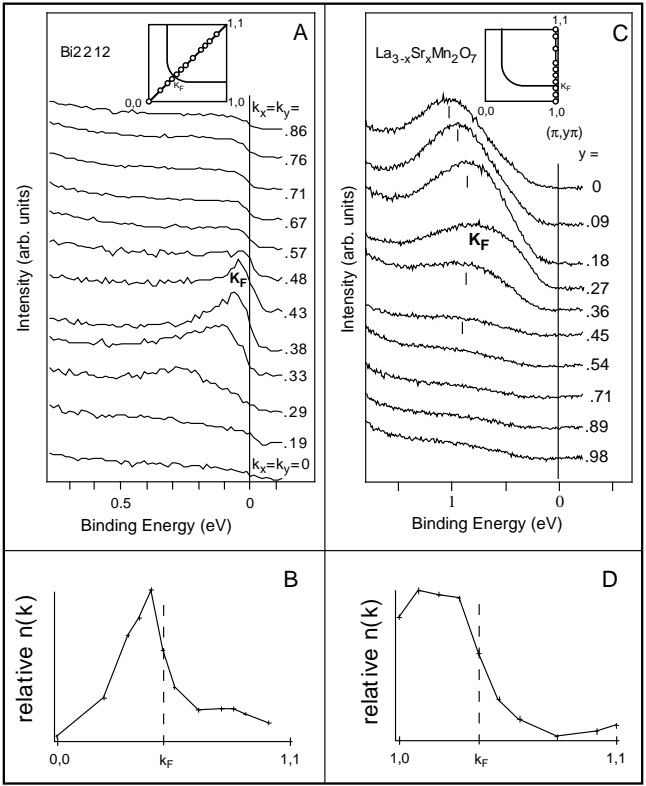}}
\vspace{0.2in}
\caption{Application of the method described in Fig.1. (A) Spectra 
along the (0,0) to $(\pi, \pi)$ cut from Bi2212.  The peak disperses 
towards the low energy side and reaches the Fermi level at 
{\it k}$_{F}$, $(0.43\pi, 0.43\pi)$.  (B) Spectra along the $(\pi,0)$ to 
$(\pi, \pi)$ cut from metallic La$_{3-x}$€Sr$_{x}$€Mn$_{2}$€O$_{7}$€.  The 
peak disperses toward the low energy side, but never reaches the Fermi 
energy.  However, it loses intensity as it crosses the position where 
the band calculation predicts the Fermi surface, showing an underlying 
Fermi surface.  (C) and (D) plots of the relative n({\it k}) for the 
data in (A) and (B), respectively, show a sudden drop around 
{\it k}$_{F}$, essentially showing the two methods give the same Fermi 
momentum {\it k}$_{F}$.}
\label{Fig2}
\end{figure}

\begin{figure}
\leavevmode \epsfxsize=\columnwidth
\centerline{\epsffile{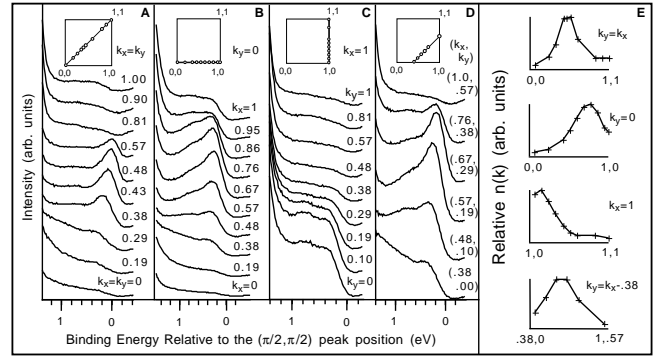}}
\vspace{0.2in}
\caption{ARPES Spectra on various cuts from 
Ca$_{2}$€Cu€O$_{2}$€Cl$_{2}$€ and n({\it k}) plots.  The insets and 
labels show where the spectra were taken in the Brillouin zone. (A) 
(0,0) to $(\pi, \pi)$ cut.  The peak disperses towards the low energy 
side and loses intensity near the $(\pi /2,\pi /2)$ point.  (B) (0,0) 
to $(\pi,0)$ cut.  The lowest energy peak shows little dispersion.  
The spectral weight initially increases and then decreases again after 
$(0.67\pi,0)$ as in the Sr$_{2}$€Cu€O$_{2}$€Cl$_{2}$€ case.  
[17][18] However, note that there is appreciable 
spectral weight at $(\pi,0)$ contrary to the 
Sr$_{2}$€Cu€O$_{2}$€Cl$_{2}$€ case.  (C) $(\pi,0)$ to $(\pi, \pi)$ 
cut.  The spectral weight drops as we move to $(\pi, \pi)$.  (D) 
Another cut (as marked in the inset) showing the n({\it k}) drop.  (E) 
relative n({\it k})'s constructed from the spectra in panels A-D. The 
relative increase of spectral weight above 0 is caused by emission 
from second order light.}
\label{Fig3}
\end{figure}

\begin{figure}
\leavevmode \epsfxsize=\columnwidth
\centerline{\epsffile{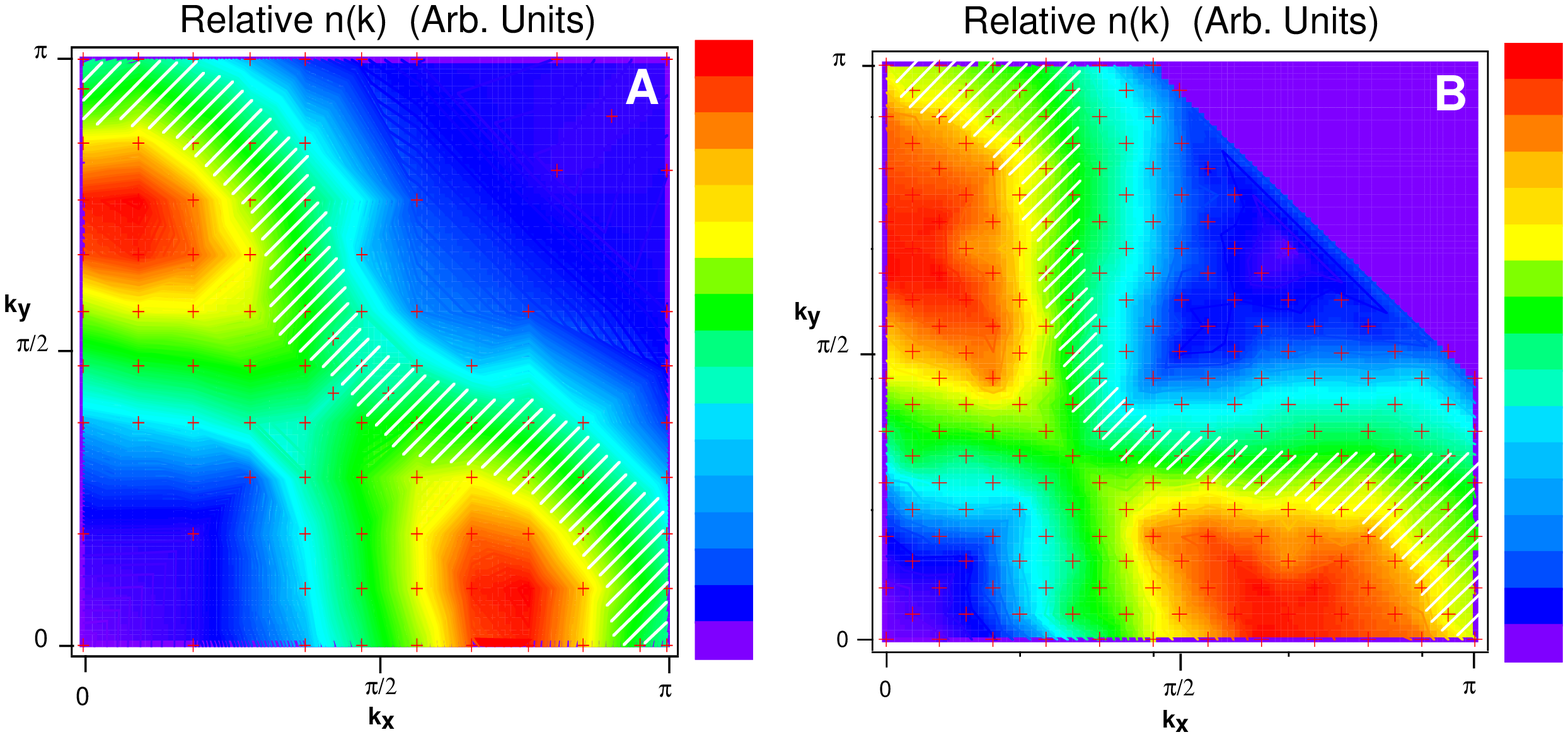}}\\ \\
\centerline{\epsffile{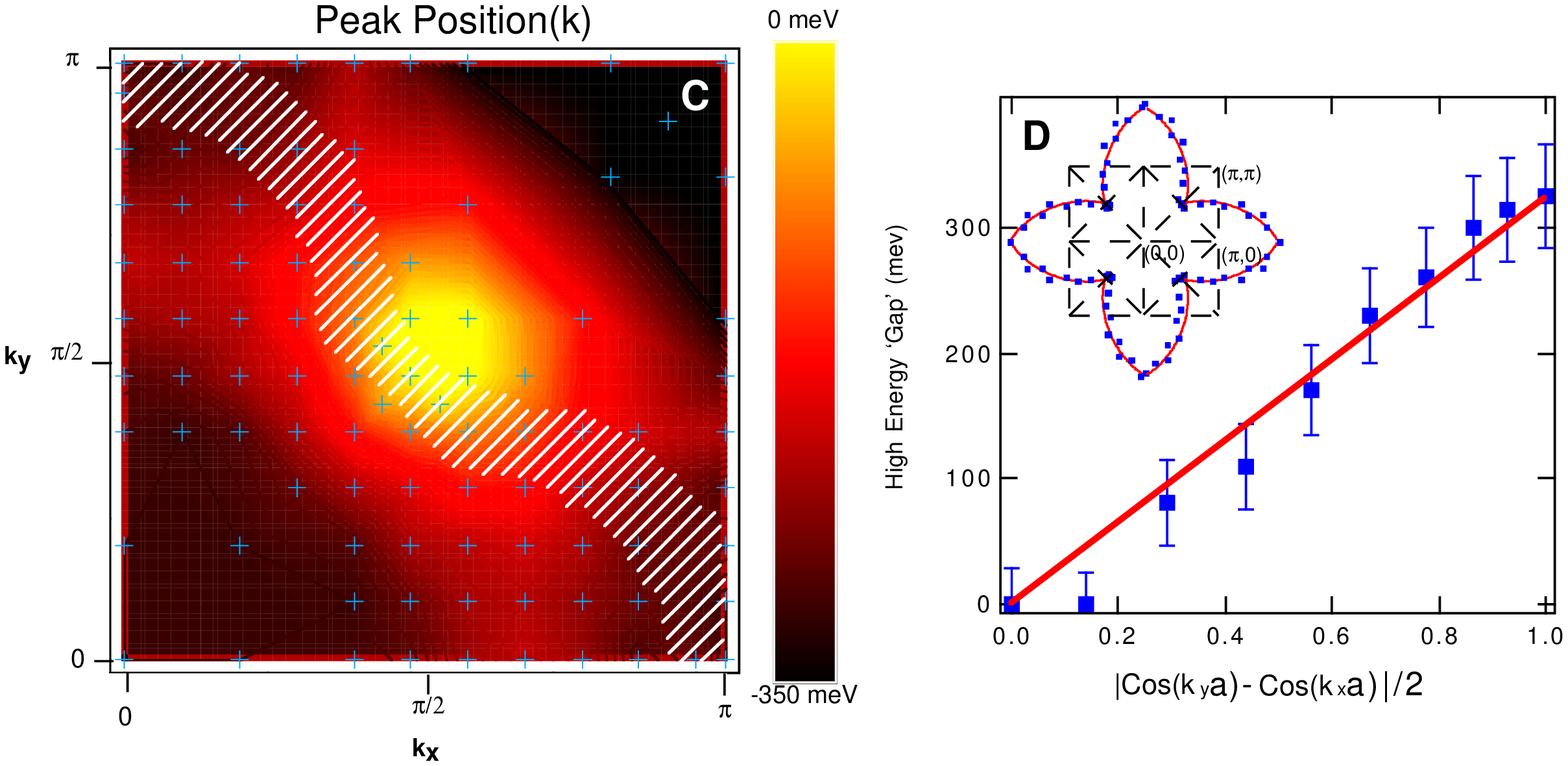}}
\vspace{0.2in}
\caption{Contour plot of the relative n({\it k}).  (A) n({\it k}) 
from the spectra shown in fig.  3.  The color scale on the right 
represent n({\it k}).  The spectra were taken only in the first octant 
of the first Brillouin zone (crosses).  The n({\it k}) plot was folded 
to better represent the remnant Fermi surface.  Note, the n({\it k}) 
drops as we cross the approximate diagonal line connecting $(0,\pi)$ 
and $(\pi, 0)$.  The hashed area represents approximately where the 
remnant Fermi surface exists.  (B) An identical plot for an optimally 
doped Bi2212 sample in the normal state.  (C) Contour plot of the 
lowest energy peak position from the spectra in fig.  3 relative to 
the $(\pi /2,\pi /2)$ peak.  The hashed area is from (A), showing the 
remnant Fermi surface.  The color scale on the right indicates the 
relative binding energy of the peak.  The peak disperses isotropically 
away from the $(\pi /2,\pi /2)$ peak position as we move away from the $(\pi 
/2,\pi /2)$ point.  (D) The 'gap' versus 
$|$cos($\it k$$_{x}$a)-cos($\it k$$_{y}$a)$|$.  The straight line shows the {\it 
d}-wave line.  The inset is a more illustrative figure of the same 
data as explained in the text. }
\label{Fig4AB}
\end{figure}

\begin{figure}
\leavevmode \epsfxsize=\columnwidth
\centerline{\epsffile{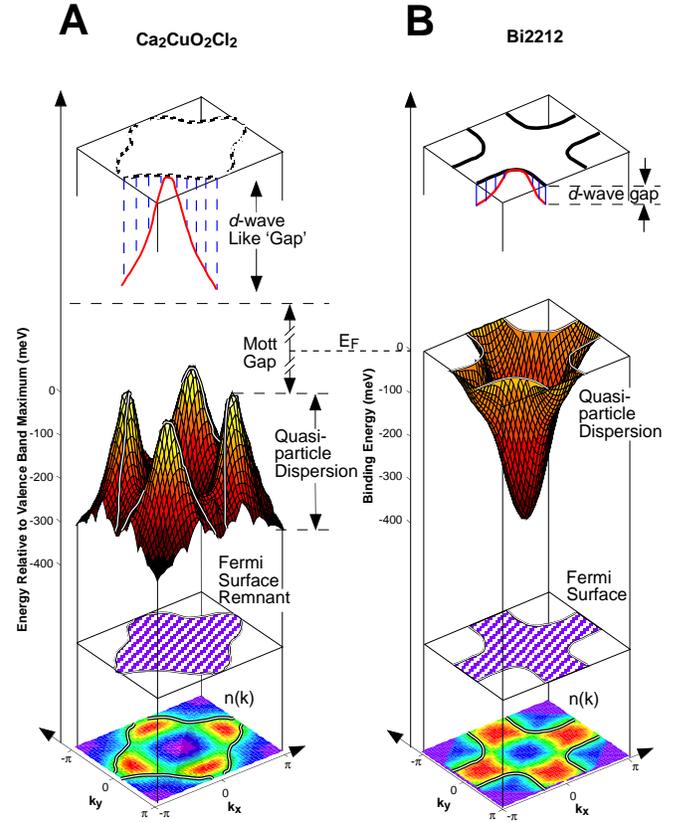}}
\vspace{0.2in}
\caption{An illustration showing the 2 experimental features 
presented in this paper on the insulator, and the similarity they show 
to a slightly overdoped Bi2212 sample.  (A) The bottom half shows the 
relative n({\it k}), and above it lies the approximate remnant Fermi 
surface derived from it.  However, there is much dispersion over the 
entire Brillouin zone, and the remnant Fermi surface is no longer an 
iso-energetic contour as can be seen by the quasiparticle dispersion 
(energy relative to the valence band maximum).  Here the remnant Fermi 
surface is shown as a black and white line running over the visible 
portions of the dispersion contour.  For clarity, a portion of the 
dispersion along the remnant Fermi surface is shown in the top half.  
Note the idea presented that the isoenergetic contour(dashed black 
line) is deformed by strong correlation to the observed red curve.  
The {\it d}-wave like 'gap' referred to in the text is the 
quasiparticle energy deviation from the dashed black line set at the 
energy of the $(\pi /2,\pi /2)$ point.  The difference between this 
'gap' and the Mott gap can now be seen clearly.  (B) For overdoped 
Bi2212, n({\it k}) defines the actual Fermi surface.  The 
quasiparticle dispersion (binding energy) shows states filled to an 
isoenergetic Fermi surface.  In the top panel one is reminded that 
below T$_{c}$, a {\it d}-wave superconducting gap opens.  This is an 
intriguing similarity between the insulator and the metal. }
\label{Fig5}
\end{figure}

\begin{figure}
\leavevmode \epsfxsize=\columnwidth
\centerline{\epsffile{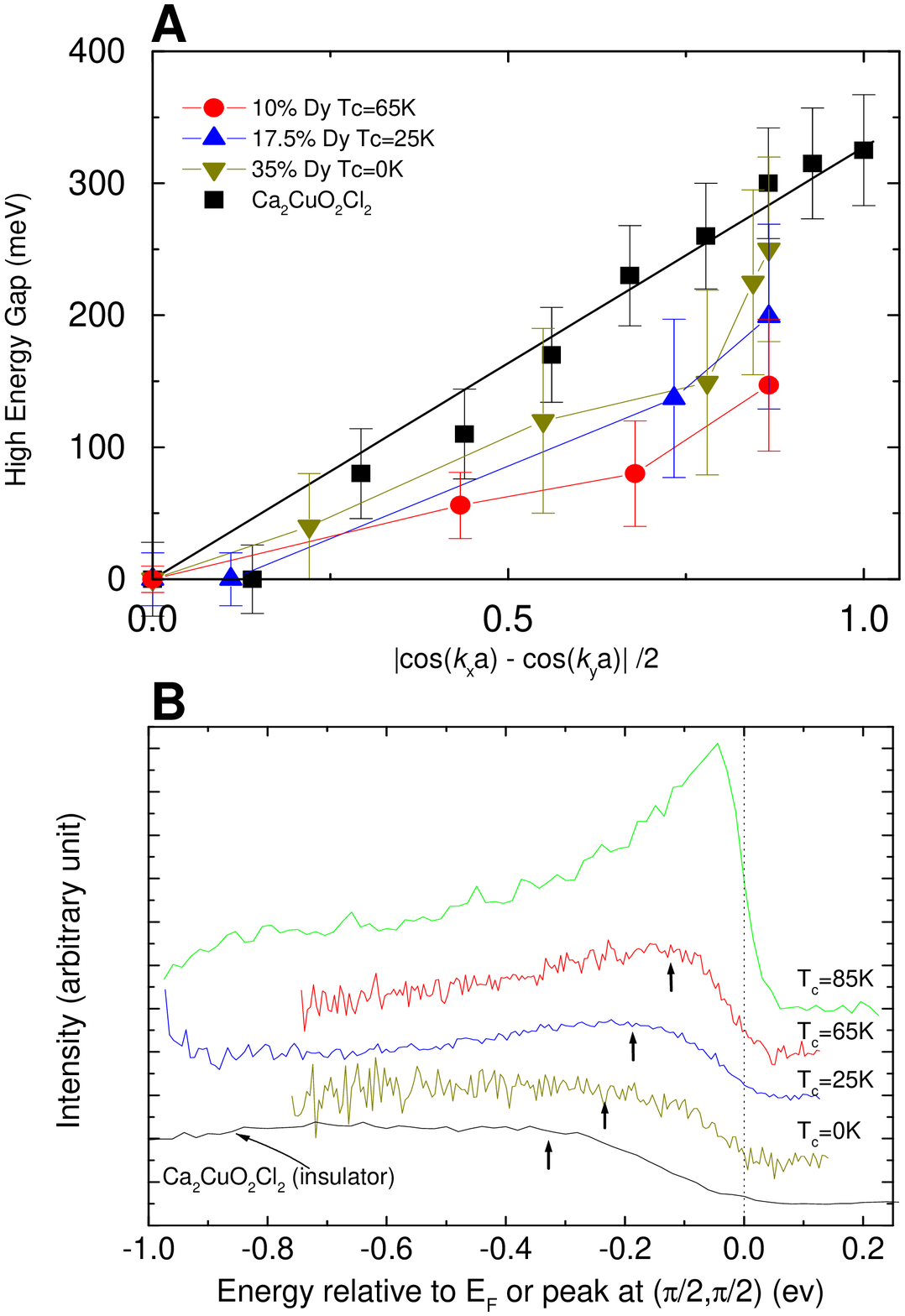}}
\vspace{0.2in}
\caption{(A) Combined {\it d}-wave plot of the data from 
Ca$_{2}$€Cu€O$_{2}$€Cl$_{2}$€ and Bi2212 with various Dy dopings.  (B) 
The spectra at $(\pi,0)$, showing the evolution of the high energy 
pseudo gap as a function of doping, as previously stressed by 
Laughlin. [12]}
\label{Fig6}
\end{figure}


\begin{references}
\bibitem{Experiment} Experiments were performed at beamline V of the 
Stanford Synchrotron Radiation Laboratory (SSRL).  
Ca$_{2}$€Cu€O$_{2}$€Cl$_{2}$€ single crystals[20] were oriented before 
the experiments by the Laue method and were cleaved in situ by 
knocking off the top posts that were glued to the samples, giving flat 
fresh surfaces suitable for ARPES experiments.  The base pressure was 
better than 5x10$^{-11}$ torr.  With the photon energies used in the 
experiments, the total energy resolution was typically 70 meV. The 
angular resolution was $\pm$1$^{o}$.  The spectra from 
Ca$_{2}$€Cu€O$_{2}$€Cl$_{2}$€ reported here were all taken at 100 K 
with 25.2 eV photons.
\bibitem{generally} As generally found in photoemission experiments, 
the change of photoemission cross-section for each sample is hard to 
calculate quantitatively.  At the qualitative level, data from 
Ca$_{2}$€Cu€O$_{2}$€Cl$_{2}$€ and Sr$_{2}$€Cu€O$_{2}$€Cl$_{2}$€ are 
consistent with each other.
\bibitem{details} The details on whether or not the dispersion near 
the {\it d}-wave node is a cusp can not yet be addressed due to the 
broadness of the observed features and the finite resolution.  The 
motivation to compare the data with the {\it d}-wave function stems 
from the fact that the insulator is related to a {\it d}-wave 
superconductor (and thus a {\it d}-wave gap) by doping.
\end{references}
\end{document}